\documentclass[twocolumn,amsmath,amssymb,pre]{revtex4}

\usepackage[]{graphicx} 
\usepackage{amsmath}

\begin{document}

\title{Free-energy functional of instantaneous correlation field in liquids:\\field-theoretic derivation of the closures}

\author{Hiroshi Frusawa}
\email{frusawa.hiroshi@kochi-tech.ac.jp}

\affiliation{Laboratory of Statistical Physics, Kochi University of Technology, Tosa-Yamada, Kochi 782-8502, Japan.}

\date{\today}

\begin{abstract}
This paper presents a unified method for formulating a field-theoretic perturbation theory that encompasses the conventional liquid state theory.
First, the free-energy functional of instantaneous correlation field is obtained from the functional-integral representation of the grand potential.
Next, we demonstrate that the instantaneous free-energy functional yields a closure relation between the correlation functions in the mean-field approximation.
Notably, the obtained closure relation covers a variety of approximate closures introduced in the liquid state theory.
\end{abstract}

%\pacs{}

\maketitle

%%%%%%%%%%%%%%%%%%%%%%%%%%%%%%%%%%%%%%%%%%%%%%%%%%%%%%
%%%%%%%%%%%%%%%%%%%%%%%%%%%%%%%%%%%%%%%%%%%%%%%%%%%%%
\section{Introduction}
Many powerful approaches to simple liquids have been developed, such as the Ornstein-Zernike integral equation theory [1-4] and the density functional [1, 5, 6] and correlation functional [7-14] theories that are based on the first and second Legendre transforms of the grand potential, respectively.
The primary goal of formulating these kinds of liquid state theory is to predict the structural and thermodynamic properties of simple liquids, considering the constituent particle arrangement or the particle-particle correlations at the atomic (or molecular) scale.

Our concern here is the closure between correlation functions [1-4],
\begin{flalign}
1+h_{\mathrm{eq}}({\bf r})=e^{-v({\bf r})+h_{\mathrm{eq}}({\bf r})-c_{\mathrm{eq}}({\bf r})+b({\bf r})},
\label{closure result}
\end{flalign}
where $h_{\mathrm{eq}}({\bf r})$ and $c_{\mathrm{eq}}({\bf r})$ denote the total correlation and direct correlation functions in equilibrium, respectively; $v({\bf r})$ is the original interaction potential in the $k_BT$-unit, and $b({\bf r})$ is referred to as the bridge function [1-4].
The Ornstein-Zernike integral equation must be combined with the relation (\ref{closure result}) using an approximate form of $b({\bf r})$ to obtain the equilibrium density-density correlation functions.

The approximations of $b({\bf r})$ in the closure relation (\ref{closure result}) have frequently resulted in thermodynamic inconsistencies, thereby limiting the physical insights and affecting the accuracy of the Ornstein-Zernike integral equation theory [1-4]. 
In addition, we have obtained some closures using uncontrolled approximations for mathematical convenience.
Accordingly, improved closure relations have been proposed using self-consistent integral equation theories [2-4].
A number of closure relations are now available for solving the Ornstein-Zernike equation, and the obtained results have been compared in detail in terms of both accuracy and thermodynamic consistency [3, 4].

Meanwhile, the correlation functional theory [7-14] investigates free energy as a functional of $h_{\mathrm{eq}}({\bf r})$, which yields the present closure (\ref{closure result}).
The correlation functional theory has thus been considered to be a promising approach to provide an integrated view on various approximations of $b({\bf r})$ because it clarifies the relation between approximate closures [1-4] and perturbative formulations based on the fugacity expansion using Mayer diagrams  [7-12].
Furthermore, a correlation functional theory that considers fluctuations and/or heterogeneities of correlation field is important, especially for glass-forming systems where out-of-equilibrium correlation functionals are to be investigated  [15-23].

It is, therefore, necessary to develop a field-theoretic perturbation theory that is able to simultaneously and seamlessly treat the meso-scale fluctuations and inhomogeneities of correlation field and the short-range correlations between particles.
Recently, we adopted the Helmholtz free energy in the canonical system to obtain the functional-integral form of the correlation functional, thereby demonstrating that the mean spherical approximation (MSA) [1-4] can be reproduced in the mean field approximation of the correlation field when using the free-energy functional in the random phase approximation (RPA) \cite{frusawa self}. 
However, it remains to be addressed to go beyond the MSA.
To this end, we first have to establish the relationship between the liquid state theory [1-14] and a field-theoretic perturbation theory.

%%%%%%%%%%%%%%%
Here we develop a field-theoretic perturbation method in the grand canonical system that encompasses the fugacity expansion used in the liquid state theory, or the conventional correlation functional theory [7-14].
In other words, our aim is twofold: (i) to formulate the free-energy functional of instantaneous correlation field and (ii) to demonstrate that a closure relation obtained in the mean-field approximation of the obtained correlation functional covers a variety of previous closures [1-4].

The remainder of this paper is organized as follows. 
In Sec. II, we define the target free-energy functional as well as instantaneous correlation field.
Section III provides the results on both the functional form and the obtained closure which will be compared with the previous ones [1-4].
In Sec. IV, we show how the present fugacity expansion is incorporated into the functional-integral representation of the grand potential as an extension of the RPA. 
In Sec. V, we outline the derivation scheme of functional integration over correlation field and its dual two-body interaction potential field.
Section VI provides concluding remarks.

%%%%%%%%%%%%%%%%%%%%%%%%%
\section{Target functional of instantaneous correlation field}
We consider $N$ particles in liquid state whose positions are specified by a set of position vectors, $\{{\bf r}_1,\cdots,{\bf r}_N\}$, thereby providing the functions, $\hat{h}({\bf r})$ and $\hat{\rho}({\bf r}')$, defined as follows:
\begin{flalign}
&\overline{\rho}^2\left\{
1+\hat{h}({\bf r})
\right\}=\hat{\rho}({\bf r}+{\bf r}')\hat{\rho}({\bf r}')
   -\hat{\rho}({\bf r}')\delta({\bf r}),
\label{instant correlation}%%%%%%%%%%%%%%%%%%%%
 \end{flalign}
and
\begin{equation}
\hat{\rho}({\bf r}')=\sum_{k=1}^N\delta({\bf r}'-{\bf r}_k),
\label{bare density}%%%%%%%%%%%%%%
\end{equation}
where the uniform density $\overline{\rho}$ is given by $\overline{\rho}=N/V$ with $V$ being the system volume.
In eq. (\ref{instant correlation}), $\hat{h}({\bf r})$ is solely dependent on the particle-particle separation vector ${\bf r}$ supposing that translational symmetry of the present system is maintained at an instant. 

Let $\Omega[v,J\equiv 0]$ denote the grand potential of liquid particles interacting via a two-body interaction potential $v({\bf r})$ in the absence of external field $J({\bf r})$.
It is noted that not only $v({\bf r})$ but also the other energies used herein are defined in the $k_BT$-unit.
We relate the above function $\hat{h}$ to an $\eta$--field via the constraint $\delta\left[\eta^2({\bf r})-1-\hat{h}({\bf r})\right]$, so that we may obtain the functional-integral representation,
\begin{flalign}
 e^{-\Omega[v,0]}
=\int D\eta\,e^{-F[\eta]},
\label{start eta}%%%%%%%%%%%%%%%%%%%%
 \end{flalign}
with the use of the identity that is associated with the following constraint (see Sec. IV.A and Appendix A for the details of introducing the functional integral):
\begin{flalign}
1=\int D\eta\,|\mathrm{det}\,\eta|
\prod_{{\bf r}}\delta\left[
\frac{\overline{\rho}^2}{2}\left\{
\eta^2({\bf r})-1-\hat{h}({\bf r})
\right\}
\right],
\label{start identity}
\end{flalign}
where the determinant contribution arising from the constant $\overline{\rho}^2$ has been absorbed into the integral measure $D\eta$.
In eq. (\ref{start eta}), $F[\eta]$ denotes the conditional free-energy functional of a given $\eta$-field, which is the target free-energy functional.

In what follows, we also rewrite $\eta^2$ as
\begin{equation}
\eta^2({\bf r})=1+h({\bf r})\geq 0,
\label{h def}%%%%%%%%%%%%
\end{equation}
where $h({\bf r})$ will be referred to as instantaneous correlation field.
Equation (\ref{h def}) reveals that the $\eta$--field is introduced to correctly reflect the following positivity  by definition of (\ref{instant correlation}) for $\hat{h}$: 
\begin{equation}
1+\hat{h}({\bf r})\geq 0.
\label{positivity}%%%%%%%%%%%%%%%%
\end{equation}
This paper focuses on whether to formulate the conditional free-energy functional of instantaneous correlation field $h({\bf r})=\eta^2({\bf r})-1$, i.e. $F[\eta=\sqrt{1+h}]$, such that it yields the
closure relation (1) to cover a variety of approximate closures used in the liquid state theory [1-4].
  
%%%%%%%%%%%%%%%%%%%%%%%%%%%%%%%%%%%%
\section{The resulting functional form}
Before proceeding to the derivation scheme, the resulting form of the target functional $F[\eta]$ is not only presented but is also assessed in comparison with the liquid state theory.

%%%%%%%%%%%%
\subsection{Conditional free-energy functional}
The present formulations are based on the premise that the instantaneous correlation field $h({\bf r})$ preserves translational invariance as mentioned above, thereby allowing the resulting form of the target functional $F[\eta]$ to be expressed as the free-energy functional per unit volume:
\begin{equation}
\frac{F[\eta]}{V}=u[\eta]-\frac{s[\eta]}{k_B},
\label{result1}
\end{equation}
where $u$ and $s$ represent the interaction energy and entropy density per unit volume, respectively.

%%%%%%%%%%%
The interaction energy density $u[\eta]$ consists of two contributions due to correlations represented by $1+h({\bf r})$ and the chemical potential $\mu$:
\begin{flalign}
u[\eta]=\frac{\overline{\rho}^2}{2}\int d{\bf r}\,\{1+h({\bf r})\}v({\bf r})-\overline{\rho}\beta\mu,
\label{result2}
\end{flalign}
where $\mu$ is multiplied by $\beta=1/k_BT$ considering that energy quantities, other than $\mu$, are defined in the $k_BT$--unit.
Meanwhile, we divide the entropy density $s$ into three parts:
\begin{equation}
s=\overline{s}+s_{\mathrm{RPA}}+\Delta s,
\end{equation}
with $\overline{s}$, $s_{\mathrm{RPA}}$ and $\Delta s$ denoting the ideal gas entropy of the uniform system, the correlation entropy obtained in the RPA and the additional contribution to the correlation entropy, respectively.

%%%%%%%%%%%%%%
As shown below, we have obtained these entropy functionals of the following forms:
\begin{flalign}
&-\frac{\overline{s}[\eta]}{k_B}=\overline{\rho}\ln\overline{\rho}-\overline{\rho},
\label{uniform s}\\%%%%%%%%%%%
&-\frac{s_{\mathrm{RPA}}[\eta]}{k_B}=
-\frac{1}{2}\int \frac{d{\bf k}}{(2\pi)^3}\,\left[\ln\{1+\overline{\rho}h({\bf k})\}-\overline{\rho}h({\bf k})\right],
\label{rpa s}\\%%%%%%%%%%%
&-\frac{\Delta s[\eta]}{k_B}=\frac{\overline{\rho}^2}{2}\int d{\bf r}\,\{1+h({\bf r})\}\{
\ln \{1+h({\bf r})\}-h({\bf r})\}\nonumber\\
&\hphantom{-\frac{\Delta s[\eta]}{k_B}=}
\qquad+\frac{\overline{\rho}^2}{2}\int d{\bf r}\,\left\{e^{h({\bf r})}-1-h({\bf r})\right\}.
\label{additional s}%%%%%%%%%%%
\end{flalign}
While the MSA can be obtained merely by discarding the additional contribution $\Delta s$, the hypernetted-chain (HNC) approximation is to take the expansion of the last term on the right hand side (rhs) of eq. (\ref{additional s}) up to the quadratic term [7-14]:
\begin{flalign}
\frac{\overline{\rho}^2}{2}\int d{\bf r}\,\left\{e^{h({\bf r})}-1-h({\bf r})\right\}
\approx
\frac{\overline{\rho}^2}{2}\int d{\bf r}\,\frac{h^2({\bf r})}{2}.
\label{hnc demo}%%%%%%%%%%%
\end{flalign}
In the next subsection, we also confirm in terms of the bridge function that the above functionals given by eqs. (\ref{result1}) to (\ref{additional s}) include the MSA as well as the HNC approximation.

%%%%%%%%%%%%%%%%
\subsection{Closure relations in the mean-field approximation}
The stationary equation of $F[\eta]$ with respect to $\eta$ is written as
\begin{flalign}
&\left.\frac{\delta F[\eta]}{\delta\eta}\right|_{\eta=\eta_m}
=2\eta_m({\bf r})\nu({\bf r})=0,\nonumber\\
&\nu({\bf r})=\left(\frac{2}{V\overline{\rho}^2}\right)\left.\frac{\delta F[\eta]}{\delta h}\right|_{h=h_m},
\label{eta stationary}
\end{flalign}
with the relation $1+h_m=\eta_m^2$ following eq. (\ref{h def}).
Equation (\ref{eta stationary}) corresponds to the mean-field equation in evaluating the grand potential $\Omega[v,0]$.
Separating the RPA contribution $\nu_{\mathrm{RPA}}$ from the remaining part $\Delta\nu$, $\nu$ is written as
\begin{equation}
\nu=\nu_{\mathrm{RPA}}+\Delta\nu,
\label{nu}
\end{equation}
where we have
\begin{flalign}
\nu_{\mathrm{RPA}}({\bf r})
&=\left(\frac{2}{\overline{\rho}^2}\right)
\left.\frac{\delta}{\delta h}\left(u-\frac{s_{\mathrm{RPA}}}{k_B}\right)\right|_{h=h_m}
=v({\bf r})+c_m({\bf r}),
\label{nu rpa}\\
\Delta\nu({\bf r})&=\left(\frac{2}{\overline{\rho}^2}\right)
\left.\frac{\delta}{\delta h}\left(\frac{\Delta s}{k_B}\right)\right|_{h=h_m}\nonumber\\
&=\ln\{1+h_m({\bf r})\}
-2h_m({\bf r})+e^{h_m({\bf r})}-1.
\label{nu delta}
\end{flalign}
In eq. (\ref{nu rpa}), $c_m({\bf r})$ can be called the direct correlation function in relation to $h_m({\bf r})$, in that $c_m({\bf r})$ is related to $h_m({\bf r})$ via the Ornstein-Zernike equation when $c_m({\bf r})$ and $h_m({\bf r})$ are regarded as the direct correlation function and total correlation function in equilibrium (i.e., $c_{\mathrm{eq}}({\bf r})$ and $h_{\mathrm{eq}}({\bf r})$), respectively.
On the one hand, eqs. (\ref{eta stationary}) to (\ref{nu delta}) read
\begin{equation}
\sqrt{1+h_m({\bf r})}\left\{
v({\bf r})+c_m({\bf r})
\right\}=0,
\label{msa general}
\end{equation}
when ignoring $\Delta\nu$.
The relation (\ref{msa general}) includes the MSA that $1+h_m({\bf r})=0\>(|{\bf r}|\leq\sigma)$ and $v({\bf r})+c_m({\bf r})=0\>(|{\bf r}|\geq\sigma)$ with $\sigma$ denoting the separation distance for exclusion such as the sphere diameter.
On the other hand, $\nu_{\mathrm{RPA}}({\bf r})+\Delta\nu({\bf r})=0$ is transformed to the typical form of the conventional closure relation (\ref{closure result}) where the subscript "eq" is replaced by "m" and the bridge function $b({\bf r})$ reads
\begin{equation} 
b({\bf r})=1+h_m({\bf r})-e^{h_m({\bf r})},
\label{bridge expression}
\end{equation}
as found from eqs. (\ref{closure result}), (\ref{nu rpa}) and (\ref{nu delta}).

While eq. (\ref{bridge expression}) is reduced to the HNC approximation in the first approximation of $e^{h_m}\approx 1+h_m$, the next approximation of $e^{h_m}\approx 1+h_m+0.5h_m^2$ transforms eq. (\ref{bridge expression}) to
\begin{flalign}
b({\bf r})\approx -0.5h^2_m({\bf r}).
\label{bridge approx}
\end{flalign}
The relation (\ref{closure result}) with the expression (\ref{bridge approx}) of $b({\bf r})$ covers various kinds of closures including the soft MSA [3, 4] and the approximations by Percus-Yevick, Verlet and Martynov-Sarkisov [1-4].

%%%%%%%%%%%%%%%%%%%%%%%%%%%%%%%%
%%%%%%%%%%%%%%%%%%%%%%%%%%%%%%%%
\section{Functional-integral representation of the grand potential}
The starting formula presented in this section is the functional-integral representation of the grand potential $\Omega[v,0]$ using four fields: one-body and two-body potential fields ($\psi$ and $\lambda$) as well as instantaneous density and correlation fields ($\rho$ and $\eta^2$).
As a result of this transformation from the configurational-integral form to the functional-integral one, we have the grand potential of a virtual system that consists of particles interacting via an imaginary two-body interaction potential with an imaginary one-body external field applied.
Hence, we also provide a field-theoretic formulation that combines the fugacity expansion and the quadratic approximation of density and one-body potential fields, in order to create a framework for evaluating the grand potential of the virtual system.
%%%%%%%%%%%%
\subsection{Introduction of functional integrals}
Multiplying the identity (\ref{start identity}) on the $\eta$--field by another identity, $1=\int D\rho\,\prod_{{\bf r}}\delta\left[\rho({\bf r})-\hat{\rho}({\bf r})\right]$, on the density field, we have
\begin{flalign}
1&=\int D\rho\prod_{{\bf r}}\delta\left[
\rho({\bf r})-\hat{\rho}({\bf r})
\right]
\nonumber\\
&
\times\int D\eta\,|\mathrm{det}\,\eta|
\prod_{{\bf r}}\delta\left[
\frac{\overline{\rho}^2}{2}\left\{
\eta^2({\bf r})-1-\hat{h}({\bf r})
\right\}
\right],
\label{identity}%%%%%%%%%%%%%%%%%%%%
 \end{flalign}
which is incorporated into the configurational-integral representation (see Appendix A) of the grand potential $\Omega[v,0]$.
The Fourier transform representation (\ref{identity}) generates two additional fields, $\psi$ and $\lambda$, which are conjugate to $\rho$ and $\eta^2$, respectively.
Furthermore, the functional integration over the $\lambda$--field is distorted by setting that
\begin{equation}
\lambda=i\lambda_{\mathrm{ref}}+w,
\label{potential division}
\end{equation}
so that we can investigate fluctuations around a reference system consisting of liquid particles that interact via a real part of reference interaction potential $\lambda_{\mathrm{ref}}({\bf r})$.
As detailed in Appendix A, the starting form of the functional integral with the integral measure $D\lambda$ replaced by $Dw$ reads
\begin{flalign}
e^{-\Omega[v,0]}&=\int D\eta\,e^{-F[\eta]}\nonumber\\
&=\iiiint
D\eta Dw D\psi D\rho
\,|\mathrm{det}\,\eta|\,e^{-F[\eta,w,\psi,\rho]},
\label{start main}%%%%%%%%%%%%
 \end{flalign}
using $F[\eta,w,\rho,\psi]$ as a functional of four fields.
We separate the $\eta$--dependent contribution $F_1[\eta,w]$ from $F[\eta,w,\rho,\psi]$:
\begin{flalign}
F[\eta,w,\psi,\rho]=F_1[\eta,w]+F_2[w,\psi,\rho],
\end{flalign}
where
\begin{flalign}
&F_1[\eta,w]=\frac{\overline{\rho}^2}{2}\iint d{\bf r}d{\bf r}'\eta^2({\bf r})\left\{
v({\bf r})-\lambda_{\mathrm{ref}}({\bf r})+iw({\bf r})
\right\},
\label{start f1}\\%%%%%%%%%
&F_2[w,\psi,\rho]
=\frac{1}{2}\iint d{\bf r}d{\bf r}'\lambda_{\mathrm{ref}}({\bf r})
\left\{\rho({\bf r}+{\bf r}')\rho({\bf r}')-\rho({\bf r}')\delta({\bf r})\right\}\nonumber\\
&\hphantom{\frac{1}{2}\iint d{\bf r}d{\bf r}'\lambda_{\mathrm{ref}}}
+\int d{\bf r}i\psi({\bf r})\rho({\bf r})+\Omega[-iw,-i\psi].
\label{start f2}%%%%%%%%%
\end{flalign}
The last term $\Omega[-iw,-i\psi]$ on the rhs of eq. (\ref{start f2}) corresponds to the grand potential of the above virtual system where particles interact via an imaginary two-body interaction potential $-iw({\bf r})$ under application of an imaginary one-body potential $-i\psi({\bf r})$.

%%%%%%%%%%%%%%%%%%%%%%
\subsection{Fugacity expansion with quadratic approximation of density and one-body potential fields}
The fugacity expansion of the grand potential $\Omega[-iw,-i\psi]$ leads to
\begin{flalign}
&\Omega[-iw,-i\psi]\nonumber\\
&\approx-\int d{\bf r}\,e^{\beta\mu+i\psi({\bf r})}
+\frac{e^{2\beta\mu}}{2}\iint d{\bf r}_1d{\bf r}_2f[\psi],
\label{main fugacity}%%%%%%%%%%%%%%%%
\end{flalign}
where the Mayer-type function,
\begin{eqnarray}
f[\psi]&=&e^{\int d{\bf r}\,i\psi(r)\hat{\rho}_2({\bf r})}\left\{
1-e^{iw({\bf r}_1-{\bf r}_2)}\right\},
\label{mayer}%%%%%%%%%%%%%%%%
\end{eqnarray}
has been introduced with the use of two particle density, $\hat{\rho}_2({\bf x})=\sum_{k=1}^2\delta({\bf r}-{\bf r}_k)$ (see Appendix B for the details).

Next we express the uniform density $\overline{\rho}$ as
\begin{flalign}
\overline{\rho}&=e^{\beta\mu-\overline{\psi}},
\label{uniform density}%%%%%%%%%
\end{flalign}
using a spatially invariant one-body potential $\overline{\psi}$, so that fluctuating fields of not only density but also one-body potential can be extracted as
\begin{flalign}
n({\bf r})&=\rho({\bf r})-\overline{\rho},\nonumber\\
\phi({\bf r})&=\psi({\bf r})-i\overline{\psi}.
\label{fluctuating fields}%%%%%%%%%%%
\end{flalign}
It is one of the essential approximations in this study to express eq. (\ref{main fugacity}) as
\begin{flalign}
&\Omega[-iw,-i\psi]-\Omega[-iw,\overline{\psi}]\nonumber\\
&\approx
\int d{\bf r}\,\left\{-i\overline{\rho}\phi({\bf r})
+\frac{\overline{\rho}}{2}\phi^2({\bf r})
\right\}
+\frac{\overline{\rho}^2}{2}\iint d{\bf r}_1d{\bf r}_2f[\phi],
\label{quadratic fugacity}%%%%%%%%%%%%%%%%
\end{flalign}
where use has been made of the following relation,
\begin{equation}
e^{2\beta\mu+\int d{\bf r}i\psi({\bf r})\hat\rho_2({\bf r})}
=\overline{\rho}^2e^{\int d{\bf r}i\phi({\bf r})\hat\rho_2({\bf r})}:
\label{mayer expansion}
\end{equation}
the quadratic expansion around $\overline{\psi}$ is performed only for the first term on the rhs of eq. (\ref{main fugacity}) while leaving the Mayer-type function $f(\phi)$ unexpanded.

We consider the conditional free-energy difference $\Delta F_2[w,\phi,n]$ between $F_2[w,\psi,\rho]$ and the spatially invariant contribution $\overline{F}[\overline{\rho}]$: 
\begin{equation}
\Delta F_2[w,\phi,n]=F_2[w,\psi,\rho]-\overline{F}[\overline{\rho}],
\label{fluctuating}
\end{equation}
where $\overline{F}[\overline{\rho}]$ is given by
\begin{flalign}
\frac{\overline{F}[\overline{\rho}]}{V}
&=\frac{\overline{\rho}^2}{2}\int d{\bf r}\lambda_{\mathrm{ref}}({\bf r})-\frac{\overline{\rho}}{2}\lambda_{\mathrm{ref}}({\bf 0})-\overline{\psi}\overline{\rho}-\overline{\rho}\nonumber\\
&=\frac{\overline{\rho}^2}{2}\int d{\bf r}\lambda_{\mathrm{ref}}({\bf r})-\frac{\overline{\rho}}{2}\lambda_{\mathrm{ref}}({\bf 0})-\overline{\rho}\beta\mu+\overline{\rho}\ln\overline{\rho}-\overline{\rho},
\label{overline f}%%%%%%%%%%
\end{flalign}
which represents the mean-field free energy of a uniform system of particles interacting via a reference interaction potential $\lambda_{\mathrm{ref}}({\bf r})$.
It follows from eqs. (\ref{start f2}), (\ref{quadratic fugacity}) and (\ref{overline f}) that eq. (\ref{fluctuating}) reads
\begin{flalign}
&\Delta F_2[w,\phi,n]=
\frac{1}{2}\iint d{\bf r}d{\bf r}'\lambda_{\mathrm{ref}}({\bf r})
n({\bf r}+{\bf r}')n({\bf r}')\nonumber\\
&\hphantom{\Delta F_2[w,\phi,n]=\frac{1}{2}\iint}
+\int d{\bf r}\,\left\{i\phi({\bf r})n({\bf r})+\frac{\overline{\rho}}{2}\phi^2({\bf r})\right\}\nonumber\\
&\hphantom{\Delta F_2[w,\phi,n]=\frac{1}{2}\iint}
+\frac{\overline{\rho}^2}{2}\iint d{\bf r}_1d{\bf r}_2\,f[\phi],
\label{delta f2}
\end{flalign}
without expanding the Mayer-type function $f(\phi)$.

We are now ready to evaluate the functional integration over the three kinds of fluctuating fields:
density $n({\bf r})$ and one-body potential $\phi({\bf r})$ as well as two-body interaction potential $w({\bf r})$.
It is found from combining eqs. (\ref{start main}) to (\ref{delta f2}) that the functional-integral representation (\ref{start main}) reads
\begin{flalign}
e^{-F[\eta]}=e^{-\overline{F}}\iiint
Dw D\phi Dn
\,|\mathrm{det}\,\eta|\,e^{-F_1[\eta,w]-\Delta F_2[w,\phi,n]},
\label{representation result}%%%%%%%%%%%%
 \end{flalign}
which will be evaluated in the next section.

%%%%%%%%%%%%%%%%%%%%%%
%%%%%%%%%%%%%%%%%%%%%%
\section{Functional integration: derivation scheme of eqs. (\ref{result1}) to (\ref{additional s})}
There are two steps in performing the functional integrals given by eq. (\ref{representation result}).
First, we evaluate the Mayer-type function $f[\phi]$ perturbatively via the functional integrations over fluctuating density and one-body potential fields, i.e. the $n$-- and $\phi$--fields. 
Next, we make use of the saddle-point approximation of the functional integral over a residual two-body interaction potential $w({\bf r})$.
It follows that the determinant term $|\mathrm{det}\,\eta|$ in eq. (\ref{representation result}) will be canceled as seen below.
%%%%%%%%%%%%%%%%
\subsection{A perturbative treatment of the Mayer-type function}
Going back to eq. (\ref{delta f2}), it is found that the usual field-theoretic perturbation method can apply to the Mayer-type function when performing the functional integrals of $e^{-\Delta F_2[w,\phi,n]}$ over $n$-- and $\phi$--fields in eq. (\ref{representation result}).
Accordingly, we have
\begin{flalign}
&e^{-\Delta F_2[w]}=\iint D\phi Dn\,e^{-\Delta F_2[w,\phi,n]}\nonumber\\
&\frac{\Delta F_2[w]}{V}=
\frac{1}{2}\int \frac{d{\bf k}}{(2\pi)^3}\,\ln\left\{1+\overline{\rho}\lambda_{\mathrm{ref}}({\bf k})\right\}
+\frac{\overline{\rho}^2}{2}\int d{\bf r}\left<f[\phi]\right>_{\phi},
\label{n phi integral}
\end{flalign}
where the perturbative treatment of the Mayer-type function $f[\phi]$ yields
\begin{flalign}
\left<f[\phi]\right>_{\phi}
&=\frac{\int D\phi\,f[\phi]\,e^{-\frac{1}{2}\iint d{\bf r}d{\bf r}'
\phi({\bf r}+{\bf r}')\gamma^{-1}({\bf r})\phi({\bf r}')}}{\int D\phi\,e^{-\frac{1}{2}\iint d{\bf r}d{\bf r}'
\phi({\bf r}+{\bf r}')\gamma^{-1}({\bf r})\phi({\bf r}')}}\nonumber\\
&=e^{-\gamma({\bf r}_1-{\bf r}_2)}\left\{
1-e^{iw({\bf r}_1-{\bf r}_2)}\right\},
\label{perturbation mayer}%%%%%%%
\end{flalign}
with a reference function defined by
\begin{flalign}
\gamma^{-1}({\bf r})=
\lambda_{\mathrm{ref}}^{-1}({\bf r})+\overline{\rho}\delta({\bf r}),
\label{gamma def}
\end{flalign}
using a shifted potential $\lambda_{\mathrm{ref}}({\bf r})$ that represents two-body interactions in a reference system (see also eq. (\ref{overline f})).
Equation (\ref{representation result}) thus reads
\begin{flalign}
e^{-F[\eta]}&=
\int Dw
\,|\mathrm{det}\,\eta|\,
e^{-F[\eta,w]}\nonumber\\
F[\eta,w]&=\overline{F}+F_1[\eta,w]+\Delta F_2[w],
\label{wonly integral}
\end{flalign}
where $\overline{F}$, $F_1[\eta,w]$ and $\Delta F_2[w]$ has been given by eqs. (\ref{overline f}), (\ref{start f1}) and (\ref{n phi integral}), respectively.

%%%%%%%%%%%%%%%%%%%%%%%%%%
\subsection{Saddle-point approximation of two-body interaction potential field}
The saddle-point approximation further divides the residual interaction potential field $w$  into two parts:
\begin{equation}
w=iw_m+u,
\end{equation}
where $u$ denotes a fluctuating potential around the mean-field potential $iw_m$ that satisfies the following stationary equation:
\begin{equation}
\left.\frac{\delta F[\eta,w]}{\delta w}
\right|_{w=iw_m}
=0.
\label{w stationary}%%%%%%
\end{equation}
We calculate the functional derivative with respect to $w$ in eq. (\ref{w stationary}) using the expressions (\ref{start f1}) and (\ref{perturbation mayer}).
Equation (\ref{w stationary}) then reads
\begin{equation}
\eta^2=1+h=e^{-\gamma-w_m}.
\label{wm solution}
\end{equation}
It follows that the second derivative is simply written as
\begin{equation}
\left.\frac{\delta^2 F[\eta,w]}{\delta w^2}
\right|_{w=iw_m}=\frac{\overline{\rho}^2}{2}\eta^2,
\end{equation}
thereby transforming eq. (\ref{wonly integral}) to 
\begin{flalign}
e^{-F[\eta]}&=\int
Du
\,|\mathrm{det}\,\eta|\,e^{-F[\eta,iw_m]-\frac{\overline{\rho}^2}{4}\iint d{\bf r}d{\bf r}'\eta^2({\bf r})u^2({\bf r})}\nonumber\\
&=e^{-F[\eta,iw_m]},\label{saddle integral}\\
F[\eta]&=F[\eta,iw_m]=\overline{F}+F_1[\eta,iw_m]+\Delta F_2[iw_m].
\label{feta general}
\end{flalign}
In eq. (\ref{saddle integral}), the $u$--integral on the rhs of the first line cancels the determinant term $|\mathrm{det}\,\eta|$ in eq. (\ref{saddle integral}).

So far, we have not specified the reference potential $\lambda_{\mathrm{ref}}$.
In this study, we set that
\begin{flalign}
-\lambda_{\mathrm{ref}}({\bf r})=c({\bf r}).
\label{lambda dcf}
\end{flalign}
The benefit of adopting this form (\ref{lambda dcf}) is that $-\gamma$ defined by eq. (\ref{gamma def}) can be identified with the instantaneous total correlation function:
\begin{flalign}
-\gamma({\bf r})
=\eta^2({\bf r})-1=h({\bf r}).
\label{gamma total}
\end{flalign}
Combining eqs. (\ref{feta general}) to (\ref{gamma total}) as well as eq. (\ref{wm solution}), we finally obtain the results given by eqs. (\ref{result1}) to (\ref{additional s}) (see Appendix C for the detailed derivation).  

%%%%%%%%%%%%%%%%%%%%%%%%%%%%%%%%%%%%%%%%%%%%%%%%%%%%%%
%%%%%%%%%%%%%%%%%%%%%%%%%%%%%%%%%%%%%%%%%%%%%%%%%%%%%%
\section{Concluding remarks}
It is noted that the relation (\ref{msa general}) of the MSA may not be derived merely from the RPA because the determinant term $|\mathrm{det}\,\eta|$ remains in the RPA.
As seen below, the MSA needs to retain the derivation process of eqs. (\ref{saddle integral}) and (\ref{feta general}) so that the present determinant term may be canceled.
It follows from eqs. (\ref{wm solution}) and (\ref{gamma total}) that the mean-field interaction potential $w_m({\bf r})$ is expressed as
\begin{flalign}
w_m({\bf r})
=h({\bf r})-\ln\{1+h({\bf r})\}.
\end{flalign}
The MSA can be validated in the approximation that
\begin{flalign}
w_m({\bf r})
\approx 0,
\end{flalign}
or its equivalent,
\begin{flalign}
e^{h({\bf r})}-1-h({\bf r})\approx 0,
\end{flalign}
while the derivation processes of eqs. (\ref{saddle integral}) and (\ref{feta general}) remain unchanged.
Thus, it is indispensable as the first approximation to perform the saddle-point approximation of the functional integral over the $w$-field.

Both of the present formulation of the grand potential and another correlation field theory \cite{frusawa self} for the canonical systems provide the same approximate relation (\ref{msa general}) of the MSA.
Nevertheless, there is a gap between the two theories.
The distinction is associated with how we cancel the determinant term $|\mathrm{det} \eta|$ that necessarily appears as long as the instantaneous correlation field $h$ is related to the $\eta$--field as $\eta^2=1+h$.
It is noted that the mean-field approximation is unable to yield eqs. (\ref{closure result}) and (\ref{bridge expression}) without introducing the $\eta$--field, and it is indispensable to introduce the $\eta$--field in both of the grand canonical and canonical theories.

The main differences between the present grand canonical and previous canonical formulations are as follows:
\begin{itemize}
\item {\itshape The Ornstein-Zernike equation}.--- The grand potential $\Omega[v,0]$ has been used in this study, so that the Ornstein-Zernike equation may be satisfied exactly.
Conversely, in the canonical systems, the Ornstein-Zernike equation is valid approximately and there actually exist correction terms \cite{white} that have been dropped in the previous formulation \cite{frusawa self}. 
\item {\itshape Fugacity expansion}.--- The above determinant term in the canonical systems is naturally canceled without the necessity of fugacity expansion, which has motivated the use of the canonical formulation \cite{frusawa self}.
As shown in eq. (\ref{saddle integral}), in contrast, the reduction of the present determinant term in the grand canonical formulation is not only due to the saddle-point approximation of the fluctuating two-body interaction term, but also due to the fugacity expansion term that is evaluated perturbatively in a field-theoretic manner.
\item {\itshape Consistency with previous closure relations}.--- Another benefit of the fugacity expansion is that it can be used to derive a kind of closure relation given by eqs. (\ref{closure result}) and (\ref{bridge expression}) that covers a variety of previous closures in addition to either the HNC approximation or the MSA [1-4], which is the central result of the grand canonical formulation.
Conversely, there is no available field-theoretic perturbation method that can reproduce the previous closures other than the MSA in the canonical systems previously considered \cite{frusawa self}.  
\end{itemize}
The above comparison thus summarizes the advantages of our functional-integral theory in the grand canonical systems that has provided the free-energy functional $F[\eta]$ of instantaneous correlation field as given by eqs. (\ref{result1}) to (\ref{additional s}).

There are two future directions of our correlation functional theory.
One direction is to consider either
fluctuating or frozen correlation fields around the mean-field correlation $\eta_m^2=1+h_m$, benefiting from the functional-integral form (4);
this allows us to investigate glass-forming systems in a field-theoretic manner [15-22].
Another direction is to provide a more elaborate form of $F[\eta]$ by incorporating higher-order terms systematically.
Such an advancement allows us not only to improve the obtained relations (1) and (20), which is consistent with the conventional closures in the liquid state theory [1-4], but also to obtain the free-energy functional including three-body correlations, which have been found to be significant in glassy systems [15-23].

In particular, it has been one of the most significant issues in the liquid state theory to improve the representation of the bridge function $b({\bf r})$.
Therefore, we would like to give more details of the latter perspective on the future studies, in terms of what kinds of improvements on the closure relation (1) are expected.
A variety of advanced evaluations beyond the saddle-point approximation remains, and it opens up the possibility of furthering the liquid state theory either to consider higher order terms for the $w$-and $\eta$-fields, or to change the reference two-body interaction potential $\lambda_{\mathrm{ref}}({\bf r})$ from the instantaneous direct correlation function $c({\bf r})$ to another function.
As a final remark, we give instances of modified evaluations for these fields of $w({\bf r})$, $\eta({\bf r})$ and $\lambda_{\mathrm{ref}}({\bf r})$:
\begin{itemize}
\item {\itshape Fluctuating two-body potential field $w({\bf r})$}.---
It is straightforward to evaluate the contribution of higher order terms beyond the saddle-point approximation.
We can easily see that the inverse power term of $\eta^2=1+h$ is created, as an additional contribution to $F[\eta]$ given by eqs. (8) to (13).
\item {\itshape Instantaneous correlation field $\eta^2({\bf r})$}.---
We have focused solely on the mean-field solution $\eta_m$ instead of taking into account fluctuations around $\eta_m$.
Prior to considering the fluctuating field, however, the solution of the mean-field equation itself has some potential for development of the liquid state theory.
First, it is interesting to see what benefits are provided by a new form of the bridge function $b({\bf r})$ when modifying the target free energy functional $F[\eta]$ due to the inclusion of higher order terms of $w$-field.
Second, eq. (15) clarifies that the mean-field solution of $1+h_m({\bf r}) = 0$ for $|r|\leq\sigma$ is also available for any approximations, other than the MSA, which can justify the conventional hybrid approximations such that the MSA and HNC approximation are applied to the short-and long-range behaviors, respectively [1, 3, 4].
\item {\itshape Reference two-body potential field $\lambda_{\mathrm{ref}}({\bf r})$}.---
In this study, the reference field is identified with the direct correlation function on an ad hoc basis.
Actually, the choice of $\lambda_{\mathrm{ref}}({\bf r})=-c({\bf r})$ successfully recovers the conventional liquid state theory, and yet it is tractable to obtain a new free-energy functional of $F[\eta]$ when $\lambda_{\mathrm{ref}}({\bf r})=h({\bf r})$ is adopted, for instance;
it follows that $\gamma({\bf r})=c({\bf r})$ as seen from Appendix C.
It remains to be investigated whether the resulting functional form of $F[\eta]$ associated with the change
of $\lambda_{\mathrm{ref}}({\bf r})$ can be of any help to the liquid state theory.
\end{itemize}

%%%%%%%%%%%%%%%%%%%%%%%%%%%%%%%%%%%%%%%%%
%%%%%%%%%%%%%%%%%%%%%%%%%%%%%%%%%%%%%%%%%
%%%%%%%%%%%%%%%%%%%%%%%%%%%%%%%%%%%%%%%%%
\appendix
%%%%%%%%%%%%%%%%
\section{Derivation of eq. (\ref{start main}) to (\ref{start f2})}
Let $\Omega[v,J]$ be the grand potential of particles interacting via a two-body interaction potential $v({\bf r}+{\bf r}')$ in the presence of external field $J({\bf r}')$.
The grand potential is given by
\begin{flalign}
e^{-\Omega[v,J]}&=\mathrm{Tr}\,e^{-H[v,J,\hat{\rho}]},\nonumber\\
\mathrm{Tr}&=\sum_{N=0}^{\infty}\frac{e^{N\beta\mu}}{N!}\int d{\bf r}_1\cdots\int d{\bf r}_N,
\label{omega def1}
\end{flalign}
where $H[v,J,\hat{\rho}]$ is defined by
\begin{flalign}
H&[v,J,\hat{\rho}]
=\frac{1}{2}\iint d{\bf r}d{\bf r}'\,v({\bf r})\left\{
\hat{\rho}({\bf r}+{\bf r}')\hat{\rho}({\bf r}')-\hat{\rho}({\bf r}')\delta({\bf r})
\right\}\nonumber\\
&\hphantom{H[v,J,\hat{\rho}]=\frac{1}{2}\iint d{\bf r}d{\bf r}'v({\bf r})}
+\int d{\bf r}J({\bf r})\hat{\rho}({\bf r}),
\label{omega def2}
\end{flalign}
using the Dirac delta density function $\hat{\rho}({\bf r})=\sum_{k=1}^N\delta({\bf r}-{\bf r}_k)$.

The Fourier transform of the identity (\ref{identity}) reads
\begin{flalign}
&1=\iint\, D\psi D\rho\,e^{-\int d{\bf r}i\psi({\bf r})\left\{\rho({\bf r})-\hat{\rho}({\bf r})\right\}}\nonumber\\
&\times\iint D\eta D\lambda\,|\mathrm{det}\,\eta|
e^{
-\frac{\overline{\rho}^2}{2}\iint d{\bf r}d{\bf r}'\left\{
\eta^2({\bf r})-1-\hat{h}({\bf r})
\right\}i\lambda({\bf r})}.
\label{identity2}%%%%%%%%%%%%%%%%%%%%
 \end{flalign}
It follows from eqs. (\ref{omega def1}) to (\ref{identity2}) that
\begin{flalign}
&e^{-\Omega[v,0]}=\mathrm{Tr}\,e^{-H[v,0,\hat{\rho}]}
\iint\, D\psi D\rho \,e^{-\int d{\bf r}i\psi({\bf r})\left\{\rho({\bf r})-\hat{\rho}({\bf r})\right\}}\nonumber\\
&\times\iint D\eta D\lambda\,|\mathrm{det}\,\eta|
e^{
-\frac{\overline{\rho}^2}{2}\iint d{\bf r}d{\bf r}'\left\{
\eta^2({\bf r})-1-\hat{h}({\bf r})
\right\}i\lambda({\bf r})}.
\label{omega appendix}%%%%%%%%%%%%%%%%%%%%
 \end{flalign}
The exponent in eq. (\ref{omega appendix}) is re-expressed as
 \begin{flalign}
&H[v,0,\hat{\rho}]+\frac{\overline{\rho}^2}{2}\iint d{\bf r}d{\bf r}'
\eta^2({\bf r})i\lambda({\bf r})\nonumber\\
&\qquad\qquad
=\frac{\overline{\rho}^2}{2}\iint d{\bf r}d{\bf r}'\eta^2({\bf r})\left\{
v({\bf r})-\lambda_{\mathrm{ref}}({\bf r})+iw({\bf r})
\right\}\nonumber\\
&\qquad\qquad
=F_1[\eta,w]
\label{reexpression1}
 \end{flalign}
and 
\begin{flalign}
&\int d{\bf r}i\psi({\bf r})\left\{\rho({\bf r})-\hat{\rho}({\bf r})\right\}+\frac{\overline{\rho}^2}{2}\iint d{\bf r}d{\bf r}'\left\{
-1-\hat{h}({\bf r})
\right\}i\lambda({\bf r})\nonumber\\
&\hphantom{\int d{\bf r}i\psi({\bf r})}
=H[-iw,-i\psi,\hat{\rho}]+H[\lambda_{\mathrm{ref}},i\psi,\rho],
\label{reexpression2}
 \end{flalign}
where use has made of both the replacements associated with the constraints given by eq. (\ref{identity}) and the potential division ($\lambda=i\lambda_{\mathrm{ref}}+w$; see also eq. (\ref{potential division})).

Combining eqs. (\ref{omega appendix}) to (\ref{reexpression2}), we have
\begin{flalign}
&e^{-\Omega[v,0]}\nonumber\\
&=\iiiint D\eta D\lambda D\psi D\rho\,|\mathrm{det}\eta|\,e^{-F_1[\eta,w]-H[\lambda_{\mathrm{ref}},i\psi,\rho]}\nonumber\\
&\hphantom{\iiiint D\rho D\psi D\eta D\lambda\,|\mathrm{det}\eta|\,}
\times\mathrm{Tr}\,e^{-H[-iw,-i\psi,\hat{\rho}]}\nonumber\\
&=\iiiint D\eta Dw D\psi D\rho\,\,|\mathrm{det}\eta|\,e^{-F_1[\eta,w]-F_2[w,\psi,\rho]},
\label{omega appendix2}%%%%%%%%%%%%%%%%%%%%
 \end{flalign}
where $F_1[\eta,w]$ has been given by eq. (\ref{reexpression1}) and $F_2[w,\psi,\rho]$ consists of two contributions:
\begin{flalign}
F_2[w,\psi,\rho]=H[\lambda_{\mathrm{ref}},i\psi,\rho]+\Omega[-iw,-i\psi].
\label{appendix f2}
 \end{flalign}

%%%%%%%%%%%%%%%%
\section{Derivation of eqs. (\ref{main fugacity}) and (\ref{mayer})}

In terms of the “Tr“ operator defined by eq. (\ref{omega def1}), the fugacity expansion is represented as
\begin{flalign}
\mathrm{Tr}\approx
1+e^{\beta\mu}+\frac{e^{2\beta\mu}}{2},
\label{tr}
\end{flalign}
considering the contributions in the range of $0\leq N\leq 2$.
The above expansion (\ref{tr}) is applied to the grand potential $\Omega[-iw,-i\psi]$ in eq. (\ref{appendix f2}), yielding
\begin{flalign}
&e^{-\Omega[-iw,-i\psi]}=1+\xi_1[\psi]+\xi_2[w,\psi]+\cdots\nonumber\\
&\xi_1[\psi]=e^{\beta\mu}\int d{\bf r}_1\,e^{i\psi({\bf r}_1)}\nonumber\\
&\xi_2[w,\psi]=\frac{e^{2\beta\mu}}{2}\iint d{\bf r}_1d{\bf r}_2\,e^{iw(|{\bf r}_1-{\bf r}_2|)+i\psi({\bf r}_1)+i\psi({\bf r}_2)}.
\end{flalign}
We further perform the following approximation,
\begin{flalign}
\Omega[-iw,-i\psi]&=-\ln\left(1+\xi_1[\psi]+\xi_2[w,\psi]+\cdots\right)\nonumber\\
&\approx-\xi_1[\psi]-\xi_2[w,\psi]+\frac{\xi_1^2[\psi]}{2}\nonumber\\
&=-\int d{\bf r}_1\,e^{\beta\mu+i\psi({\bf r}_1)}
+\frac{e^{2\beta\mu}}{2}\iint d{\bf r}_1d{\bf r}_2f[\psi]\nonumber\\
f[\psi]&=e^{i\psi({\bf r}_1)+i\psi({\bf r}_2)}\left\{
1-e^{iw({\bf r}_1-{\bf r}_2)}\right\},
\label{h2_xi}
\end{flalign}
so that the Mayer-type function $f[\psi]$ may appear in evaluating the interaction energy of $\Omega[-iw,-i\psi]$.

%%%%%%%%%%%%%%
\section{Transforming eqs. (\ref{saddle integral}) to (\ref{gamma total}) into the main results (\ref{result1}) to (\ref{additional s})}
The relation (\ref{gamma def}) implies that
\begin{flalign}
&\iint d{\bf r}_1d{\bf r}_2\,\lambda_{\mathrm{ref}}({\bf r}_0-{\bf r}_1)\gamma^{-1}({\bf r}_1-{\bf r}_2)\gamma({\bf r}_2-{\bf r}_3)\nonumber\\
&=
\iint d{\bf r}_1d{\bf r}_2\,\lambda_{\mathrm{ref}}({\bf r}_0-{\bf r}_1)\lambda_{\mathrm{ref}}^{-1}({\bf r}_1-{\bf r}_2)\gamma({\bf r}_2-{\bf r}_3)\nonumber\\
&+\iint d{\bf r}_1d{\bf r}_2\,\overline{\rho}\lambda_{\mathrm{ref}}({\bf r}_0-{\bf r}_1)\delta({\bf r}_1-{\bf r}_2)\gamma({\bf r}_2-{\bf r}_3),
\end{flalign}
which reads
\begin{flalign}
-\gamma({\bf r}-{\bf r}')=-\lambda_{\mathrm{ref}}({\bf r}-{\bf r}')
+\int d{\bf r}"\,\overline{\rho}\lambda_{\mathrm{ref}}({\bf r}-{\bf r}")\gamma({\bf r}"-{\bf r}'),
\end{flalign}
thereby validating eq. (\ref{gamma total}) when $\lambda_{\mathrm{ref}}({\bf r})=-c({\bf r})$.
Considering that
\begin{flalign}
1-\overline{\rho}c({\bf k})=\frac{1}{1+\overline{\rho}h({\bf k})},
\end{flalign}
we obtain $F[\eta]=F_1[\eta,iw_m]+F_2[iw_m]$ where
\begin{flalign}
&\frac{F_1[\eta,iw_m]}{V}\nonumber\\
&=\frac{\overline{\rho}^2}{2}\int d{\bf r}\{1+h({\bf r})\}\left\{
v({\bf r})+c({\bf r})+\ln\{1+h({\bf r})\}-h({\bf r})\right\}\label{appendix f1wm},\\
&\frac{F_2[iw_m]}{V}
=\frac{\overline{F}}{V}-\frac{1}{2}\int \frac{d{\bf k}}{(2\pi)^3}\,\ln\left\{1+\overline{\rho}h({\bf k})\right\}\nonumber\\
&\hphantom{\frac{F_2[iw_m]}{V}
=\frac{\overline{F}}{V}}
\qquad+\frac{\overline{\rho}^2}{2}\int d{\bf r}\left\{
e^{h({\bf r})}-1-h({\bf r})
\right\}.
\end{flalign}
Furthermore, we have
\begin{flalign}
&\frac{\overline{F}}{V}+\frac{\overline{\rho}^2}{2}\int d{\bf r}\{1+h({\bf r})\}\left\{
v({\bf r})+c({\bf r})\right\}\nonumber\\
&\qquad\qquad\qquad=u[\eta]-\frac{\overline{s}[\eta]}{k_B}+\frac{1}{2}\int \frac{d{\bf k}}{(2\pi)^3}\,\overline{\rho}h({\bf k})
\end{flalign}
because of
\begin{flalign}
&\frac{\overline{\rho}}{2}\left[c({\bf 0})+
\overline{\rho}\int d{\bf r}\,h({\bf r})c({\bf r})
\right]
=\frac{\overline{\rho}}{2}h({\bf 0})=\frac{1}{2}\int \frac{d{\bf k}}{(2\pi)^3}\,\overline{\rho}h({\bf k}),
\label{appendix zero separation}
\end{flalign}
using the Ornstein-Zernike equation at zero separation.
It is found from eqs. (\ref{appendix f1wm}) to (\ref{appendix zero separation}) that eqs. (\ref{saddle integral}) to (\ref{gamma total}) can be transformed to the main results (\ref{result1}) to (\ref{additional s}).

\end{document}